# Forensics Acquisition and Analysis of instant messaging and VoIP applications

Christos Sgaras[1], M-Tahar Kechadi[2] and Nhien-An Le-Khac[2]

[2]School of Computer Science & Informatics, University College Dublin
Belfield, Dublin 4, Ireland
`{tahar.kechadi, an.lekhac}@ucd.ie`

**Abstract.** The advent of the Internet has significantly transformed the daily activities of millions of people, with one of them being the way people communicate where Instant Messaging (IM) and Voice over IP (VoIP) communications have become prevalent. Although IM applications are ubiquitous communication tools nowadays, it was observed that the relevant research on the topic of evidence collection from IM services was limited. The reason is an IM can serve as a very useful yet very dangerous platform for the victim and the suspect to communicate. Indeed, the increased use of Instant Messengers on smart phones has turned to be the goldmine for mobile and computer forensic experts. Traces and Evidence left by applications can be held on smart phones and retrieving those potential evidences with right forensic technique is strongly required. Recently, most research on IM forensics focus on applications such as WhatsApp, Viber and Skype. However, in the literature, there are very few forensic analysis and comparison related to IM applications such as WhatsApp, Viber and Skype and Tango on both iOS and Android platforms, even though the total users of this application already exceeded 1 billion. Therefore, in this paper we present forensic acquisition and analysis of these four IMs and VoIPs for both iOS and Android platforms. We try to answer on how evidence can be collected when IM communications are used. We also define taxonomy of target artefacts in order to guide and structure the subsequent forensic analysis. Finally, a review of the information that can become available via the IM vendor was conducted. The achieved results of this research provided elaborative answers on the types of artifacts that can be identified by these IM and VoIP applications. We compare moreover the forensics analysis of these popular applications: WhatApp, Skype, Viber and Tango.

**Keywords:** VoIP; forensic acquisition analysis and comparison; WhatsApp; Skype; Viber; Tango

## 1 Introduction

One of the daily activities that has seen significant changes due to service introduced via the Internet is the social networking and communication amongst people [1][2][4]. Various Internet based communication services have been introduced that offer di-verse methods of communication such as instant messaging, audio, video, file exchange and image sharing. The term "*Instant Messaging application*" or "*IM applications*" will be used throughout this text in order to refer to this category of Internet based communication services.

However, it was not only the legitimate activities that shifted to Internet based services. Eventually, criminal activities started being facilitated or taking place over the Internet [5] and criminals started using IM applications to communicate, either with potential victims [6] or amongst themselves to avoid interception [7]. It becomes obvious that, due to their popularity, IM applications have the potential of being a rich source of evidential value in criminal investigations. Moreover, for most of these applications, the type of information that can be collected in the context of an investigation can span beyond text messages. Europol has identified the threat of misused IM communications by criminals to facilitate their illegal activities due to the fact that it is harder to monitor or regulate these services [8].

Despite the wide prevalence of IM applications, there has been limited published research towards the assessment of their evidential value in criminal investigations. Moreover, the existing published research has not addressed all the popular OS platforms and IM applications, either due to the fact that they did not all exist at the time the research was conducted or the specific IM application was not that prevalent at that point to attract the attention of deep investigations.

As a result, this acted as the main motivation in proceeding with this specific problem since any solution to it would address the gap in the existing published research and contribute to its further progress. In this paper, we present forensic acquisition and analysis of four VoIPs: WhatsApp, Skype, Viber and Tango for both iOS and Android platforms. They are the most popular IM apps as their total users already exceeded 1 billion [23][24][26]. Indeed, to the best of our knowledge, there is no published research regarding forensic artefacts of Tango as well as a comparison of forensics analysis of these four IMs. Indeed, most of the published research has been focusing on Android devices whereas iOS based devices have not been extensively examined. In this paper we aim to answer how evidence can be collected when these IMs' communications are used. In order to provide answers to this problem, the idea is to select a number of IMs services and operating systems based on current popularity and conduct test communications for further analysis. Furthermore, a taxonomy of target artefacts was defined in order to guide and structure subsequent forensic analysis. Additionally to the forensic analysis, alternative sources of evidence were examined such as the possibility to clone IM session and perform communication interception. Finally, a review of the information that can become available via the vendor was conducted. Based on this approach, experimental tests were also conducted in order to identify potential forensic artefacts for these IM applications.

The rest of this paper is structured as follows: Section 2 looks at the literature survey and relevant work that has been conducted in the field. Section 3 describes the acquisition techniques in order to address the problem as identified in this section. Section 4 presents the evaluation and discussion of the outcomes of applying the adopted approach to solve the problem. Finally, Section 5 summarises some concluding remarks and discusses some future work.

## 2      Background

Although IM and VoIP applications were not very widespread in 2006, Simon & Slay [9] had already identified that the use of VoIP communications poses new challenges to law enforcement authorities. Specifically, they pointed out that new methods for collecting evidence were needed due to the fact the VoIP communications are based on a decentralised data network and can be easily encrypted. In the same paper, the authors described the challenges of performing call interception due to VoIP architecture of non-carrier solutions with a special reference to Skype, which was one of the few VoIP applications at that time. They present a potential real life scenario where criminals could avoid the traditional PSTN network and communicate via Skype in order to avoid interceptions and achieve the necessary obscurity required to remain undetected [10]. Moreover, they referred to the immaturity of legislation regarding the regulation of non-carrier VoIP and the complexity of introducing relevant legislation at an international level. Finally, they proposed memory forensics as a potential direction in retrieving volatile evidence related to running VoIP software, although they do not present a concrete methodology since their research was still incomplete at that point.

Kiley et al [11] conducted some research on IM forensic artefacts and they also pointed out that IM is being exploited by criminals due to its popularity and privacy features. However, their research focused on, what they named, volatile instant messaging which described the IM services operating via a web interface, without requiring a fat client. They analysed four popular web-based IM services on Windows desktop environments and concluded that it is possible to retrieve forensic artefacts via the browser cache files and the Windows page files. The identified forensic artefacts included communication timelines, usernames, contact names and snippets of conversation. However, the entire conversation was never possible to be retrieved. The authors concluded their paper by presenting an investigative framework for addressing volatile messaging, which consists of three phases: recognition, formulation and search.

Simon & Slay [12] continued their work on investigative techniques for VoIP technologies by conducting experiments in order to identify traces left by Skype in the physical memory. They drew their inspiration from similar research on the operating system recovery level information from the physical memory and expanded the approach to application level information. To better structure the objectives of their proposed investigative approach, the authors defined a number of data type categories that could be identified in the context of an investigation: Communication Content, Contacts, Communication History, Passwords and Encryption Keys. In order to capture the memory of the system under investigation, they leveraged virtualisation and the inherent functionalities of memory extraction. They concluded that it is possible to retrieve useful information about the use of Skype via the physical memory and specifically: information about the existence of the Skype process, the password and the contact list of the Skype account that was used. However, it was not possible to retrieve any encryption keys although it is known that Skype uses encryption.

Vidas et al [13] conducted work towards the definition of a general methodology for collecting data on Android devices. Although not directly relevant to the IM in-vestigation issue, their work provides useful information that assisted in developing our approach. More precisely, the authors have provided a comprehensive overview of an Android device and proposed specific data collection objectives and processes that were taken into account in our experiments, which are described in Section 4.

Alghafli et al [14] gave guidelines on the digital forensic capabilities in smartphones where they considered Skype as a source of evidence. They also referred to VoIP applications being used to communicate without leaving logs in the traditional phone functions of the smartphones.

Carpene [15] and Tso et al [16] approached the evidence extraction for iOS based devices from a different perspective, which does not require the actual seizure of the device. Their approach was to leverage the backup files created via the iTunes application, the PC companion software for managing iOS devices. In [15] the author took a more generic approach in an attempt to create taxonomy for all potential evidence that can be extracted from the iTunes backup file. Skype is one of the applications that were included in this taxonomy and the potential evidence identified is limited call history and limited contact data. The same approach was followed in [16] to investigate the iTunes backup files. Nevertheless, their research focused in identified forensic artefacts of five popular IM devices: Facebook, Skype, Viber, Windows Live Messenger, and WhatsApp Messenger. They concluded that it is possible to retrieve the content of IM communications via the backup files. Since three of these IM applications are included in analysis, their outcome will be taken into consideration.

Schrittwieser et al [17] conducted a security assessment of nine popular IM and VoIP applications, including Viber, WhatsApp and Tango. Although their approach is stemming from a vulnerability assessment perspective, the outcome could also be valuable in a law enforcement investigation context. For instance, their results about weak authentication mechanisms could be used in order to perform session cloning and intercept the communication, wherever the applicable legislation allows for such activities by law enforcement authorities.

Chu et al [18] examined the possibility of retrieving Viber communication content via the Random Access Memory (RAM) in Android devices. They concluded that it is possible to retrieve partial evidence via the RAM and, furthermore, that the evidence is present even after resetting the device.

Mahajan et al [19] conducted forensic analysis for both Viber and WhatsApp on Android devices, using forensic acquisition equipment to perform the file system extraction of the smartphones. Their research concluded that for both IM applications it is possible to retrieve useful information about the user's activities such as communication content, communication history and the contact list.

We also want to validate what has been concluded in [19] and go beyond by expanding it to issues that they have not been addressed such as the recovery of deleted messages or factory reset phones. Moreover, we will expand further to more platforms and more IM applications. As presented above, IM and VoIP applications are in the rise within the last few years and most likely will remain popular in the future. The shift from traditional communications (i.e. over PSTN or GSM networks) to IM and VoIP services calls requires new ways of collecting evidence.

The existing literature has identified the issue of identifying and monitoring VoIP traffic [9][10][20] and some research has been conducted towards the retrieval of evidence from IM and VoIP services [11][12][15][16][19]. Nevertheless, there are gaps in the existing research and developments on this field, which have yet to be addressed. For instance, the investigating forensic artefact of Tango, which is popular amongst Android users or the desktop version of Viber which was only released in May 2013 is highly required. Moreover, very little is known about alternative investigation techniques which is often used by law enforcement for Skype investigations. Indeed, there is no research that compares forensics analysis of four applications: WhatsApp, Skype, Viber and Tango. Finally, most of the published research has been focusing on Android devices whereas iOS based devices have not been extensively examined.

These open issues formulate the problem statement of this work, which can be summarised in the following questions:

- What types of forensic artefacts can be found in the most popular IM services?
- How and where can evidence be collected from the most popular IM services?
- What alternative sources exist in order to capture evidence from IM services?

In order to address these questions we follow the approach that is described in the following Section.

## 3  Acquisition Techniques

The first step is to set-up an investigation environment for various mobile devices in with WhatsApp, Skype, Viber and Tango are installed. Following the environment setup and the definition of the list of target artefacts, the next steps are dedicated to the whole investigation itself; from the data collection to the extraction of evidence (artefacts in this case). We perform the forensic analysis on the data. This approach implies that a device has been seized and therefore we can conduct a forensic analysis on it in order to extract evidence in a post mortem fashion.

### 3.1  Forensics Analysis

The objective of this analysis is to identify the artefacts stored by each IM application in the file system of every seized device. The following questions are usually expected to be answered during this analysis.

- What data is generated and stored on the device for each of the used IM functionality?
- Where is this data stored on the file system?
- In what format is the data stored?
- How can the data be retrieved, accessed and analysed?

For the data extraction and analysis from the devices, we use specialised mobile forensic tools:

- Cellebrite UFED Touch Ultimate  - data extraction / acquisition – with the following extraction modes:

- Logical extraction: Quick extraction of target data (e.g. sms, emails, IM chats) performed at the OS level.
- File system extraction: In depth extraction of the entire file system of the device.
- Cellebrite UFED Physical Analyzer  - data analysis

Following the data extraction, we use SQLiteStudio as a main tool for opening and parsing the SQLite databases that are mainly used for storing data in IM applications. Further we use other software applications for opening image, audio and video files that have been identified.

### 3.2  Taxonomy of target artefacts

In this subsection, we describe taxonomy of target atefacts that we use in our forensics analysis.

**Table 1.** Target artefacts

| Target Artefacts | Definition |
| --- | --- |
| Installation Data | Data related to the installation of an IM client on a specific device. It can be very useful in the initial phase of an investigation, as it can lead to further queries related to IM data. |
| Traffic data | According to the European Data Protection Supervisor [27], *traffic data are data processed for the purpose of the conveyance of a communication on an electronic communications network. According to the means of communication used, the data needed to convey the communication will vary, but may typically include contact details, time and location data.* Therefore it is crucial for forensic analysis. |
| Content data | The actual content of a communication, which can be text, audio, video, or any other format of the data. In the context of this study, we do not categorise attachments or exchanged files as content data but we establish category for them. |
| User profile data | Information related to the profile of the IM user such as name, surname, birthdate, gender, picture, address, phone number and email. |
| User authentication data | Data that is used to authenticate the user to a service or an application such as a password, session key, etc. |
| Contact database | The list of contacts associated to the IM user. |
| Attachments/Files exchanged | Data files that were exchanged via a file transfer functionality. |
| Location Data | According to the UK Information Commission Officer [28], *location data means any data processed in an electronic communications network or by an electronic communications service that indicates the geographical position of the terminal equipment of a user of a public.* |

### 3.3 Discussion

The approach adopted in order to conduct this study is influenced by Simon & Slay [12] with regards to the definition of data categories of expected evidence. Moreover, the adopted approach for the data acquisition from the mobile devices is based on the same technique used in [19]. Nevertheless, the adopted approach differs from these previous studies in the following:

- It covers both iOS and Android platforms. Previous work mainly focused either on Android devices [19] or the backup files of iOS devices [15][16], but not both at the same time.
- It covers the four most popular IM applications; in total more than 1 billion users, on the two most popular mobile platforms, covering 63% of the market share as described earlier. Overall, the scope of the examined IM and mobile OS platforms should cover the vast majority of IM based communications via mobile devices.
- With the exception to the study presented in [19], none of the previous work was based on the file system forensic analysis of data images acquired by mobile devices.
- Although mostly focusing on the mobile file system forensic analysis, it addresses the IM communications from file system forensic analysis. Previous work was focused only on the technical forensic analysis of the file system [19], backup files [15][16] or the physical memory [12][18].

## 4 Description of results and Analysis

### 4.1 Test environment

In order to examine the four IM applications and answer the questions posed in the problem statement, a testing environment is setup, based on two different mobile devices with all the four IM applications installed (Fig.1). The communications between the two devices was conducted for a week using all features of IM applications.

We used the iOS version 6.1.3 and the Android version 2.3.5 (Gingerbread) during the testing. Although the Android operating system is characterised by considerable fragmentation in terms of versions, it can be assumed that the results of the tests can be extended to all current versions of Android as it was also concluded in [19]. It must be noted that both devices were neither jail-broken nor rooted. In the case of a jail-broken or rooted device, the test would result in at least the same amount of evidence, if not more due to the more uncontrolled environment in a jail-broken or rooted device. Moreover, the devices used were not locked with a passcode, although the equipment we used was able to retrieve almost the same data when the devices were locked. Nevertheless, the passcode recovery or cracking of the devices is out of the scope of this study.

## 4.2 iOS Forensic Analysis

**Logical Extraction Analysis.** The Logical Extraction of information from the iPhone did not produce any IM related results. The analysed data from UFED Physical Analyser do not include any explicit information related to the four IM applications under investigation.

**Filesystem Extraction Analysis.** On the other hand, when a File System Extraction was performed with UFED, the UFED Physical Analyzer was able to identify relevant information for Skype, Viber, Tango and WhatsApp (Table 2).

**Manual Filesystem Analysis.** In order to further investigate for potential file system artefacts generated by facts, it was required to proceed with a manual analysis of the file system. The most valuable information in the Skype user folder is found at the following locations: main.db.EMBEDDED that contains the profile pictures of all the Skype contacts and main.db that contains information about the Skype account, list of contacts that participated in a call, list of Skype contacts, list of conversations, messages and SMS, list of transferred files from and to the user, list of voicemails sent to the user and list of video calls.

**Table 2.** iOS Filesystem Extraction analysis

| Target artefacts | Skype | Viber | Tango | WhatsApp |
|---|---|---|---|---|
| Installation Data | Yes | Yes | Yes | Yes |
| Traffic data | Call/Chat history | Call/Chat history | No | Chat history |
| Content data | Chat/Image | Chat/Image | Image | Chat/Image |
| User profile data | Yes | No | Yes | No |
| Contact database | Yes | No | | No |

The most important file of Viber is Contacts.data that is at the core of the application and comprises the main data repository where valuable information can be found such as a list of all contacts; list of all the attachments exchanged, including pictures, video, stickers and custom locations; list of unique conversation; list of phone numbers; list of recent calls and their duration; data related to the sticker icons and packages; the latitude and longitude of each message that was sent with the location service enabled; list of messages exchanged.

The first remark that was made when researching the Tango application in the iOS device was the fact that the application is registered with the name 'sgiggle' instead of tango in the file system. Secondly, it has to be noted that Tango's application folder 'sgiggle' was the only one of the IM applications in scope that was hidden in the file system. While analysing the files included in Tango's application folder, it was initially noticed that mainly comprises of SQLite .db files like all the other examined applications. However, when attempting to parse the database files, it was noticed that the content of the databases are not in clear text. For instance, the database file tc.db appears to be the one storing the communication content in Tango. When examining the data of the table 'messages', the content of the fields 'conv_id' and 'payload' appear to be unintelligible which leads to the conclusion that they are stored in an encrypted form. The same kind of encryption was encountered with all other database files in the Tango application folder which made the further analysis of the data not feasible in the context of this research, as it would require extensive cryptanalysis attacks.

The manual file system analysis identified that the WhatsApp application files reside in an iOS folder in the /var/mobile/Applications/ directory. The application files are contained in two subfolders 'Documents' and 'Library', with the first one storing all the database and iOS plist settings files while the latter contained media files such as images and videos. The subfolder Library/Media contained the following subfolders: <contact_number>@s.whatsapp.net that contains all the media files exchanged in conversations with the specific contact; profile that contains thumbnail images of the profile pictures of all contacts. Nevertheless, the database files storing the activity and contact information for WhatsApp reside in the Documents folder. Specifically, the communication activity is stored in the database called ChatStorage.sqlite and the WhatsApp contact information is stored in the Contacts.sqlite database file.

### 4.3 Android Forensic Analysis

**Logical Extraction Analysis.** Contrary to the results retrieved for iOS via the UFED Logical Extraction analysis, the equivalent results for Android are much more extensive with a considerable amount of IM related information been identified. However, the logical extraction analysis of the Android device did not produce any information regarding Viber. Besides, the only related information to Tango was a record in the 'User Accounts' section of an account titled "Sync Tango friends". Similarly to Skype, information related to WhatsApp includes all the chat history and its content. Moreover, the WhatsApp chat content also includes the attached files (Table 3).

**Table 3.** Android Logical Extraction analysis

| Target artefacts | Skype | Viber | Tango | WhatsApp |
|---|---|---|---|---|
| Traffic data | Call/Chat history | N/A | N/A | Chat history |
| Content data | Chat/Image/Video | N/A | N/A | Chat/Image |
| User profile data | Yes | N/A | N/A | No |
| Contact database | Yes | N/A | N/A | No |

**Filesystem Extraction Analysis** of the Android device did not produce significantly different results compared to the Logical Extraction analysis described in section 4.2. For Skype, File System Extraction analysis provides more information contained in the chat content. Specifically, while the attachments were missing in the Logical Extraction analysis, most of them were embedded in the conversations retrieved by the File System Extraction analysis. Nevertheless, this information can also be retrieved vie the Logical Extraction analysis under the Videos section.

Similarly, the only possibility of retrieving Viber's data was by searching for "viber" in the Images and Videos section which returned the image and video files.

Nevertheless, it only provided an initial identification of related data without providing a lot of context.

For Tango, additionally to the information retrieved by the Logical Extraction Analysis, the File System Extraction Analysis revealed an entry of the Tango application in the 'Installed Applications' section of the UFED Physical Analyzer. Finally, for WhatsApp, this analysis does not produce significantly different results compared to the Logical Extraction analysis described above. The same data was identified with regards to WhatsApp as the ones described with the Logical Extraction analysis.

**Manual Filesystem Analysis.** With the exception of the location in the file system, the Android folder structure of Skype application data is similar to the one described for iOS. As a result, the main bulk of the information is stored in the main.db SQLite database which is located in the Skype user folder. The Skype user folder in an Android device is located at the following path /data/data/com.skype.raider/files/<username>. Apart from the information contained in the main.db database and already analysed in the iOS section, the following folders can be of interest in the Android filesystem: media that contains media files that have been exchanged via Skype and /mnt/sdcard/Android/data/com.skype.raider/cache that is located in the SD card of the device and contains cached media files that were sent by the Skype user.

Since the analysis performed by the UFED Physical Analyzer did not yield any results related to Viber, the manual file system analysis is the only possible way of retrieving evidence coming from IM communications over Viber. Contrary to Skype where the databases between iOS and Android were almost identical, the situation with Viber was identified to be considerably different between the two operating systems. In the case of Android, the information is mainly stored in the following locations:

- Folder /mnt/sdcard/viber that locates in the SD card of the device and contains the following subfolders which might be of interest in an investigation: User photos- profile photos of all the Viber contacts, Viber Images- Image files exchanged via Viber, Viber Video-Video files exchanged via Viber
- Folder /data/data/com.viber.voip: Located in the device's internal memory and containing the following database files

When analysing the two database files, we notice that the viber_message is used for storing information related to text communications whereas the viber_data for the rest such as the call log and the phonebook.

The manual file system analysis in the Android file system resulted in identical conclusions as the one described in the iOS section. All the Tango database files contained encrypted content and TangoCache.db was the one in cleartext pointing to exchanged media files. The difference compared to the iOS analysis was the fact that the folder TCStorageManagerMediaCache and the contained media files were available locally.

Mahajan et al had identified the artefacts produced in WhatsApp in their research [19]. However, they do not include the exact structure of the WhatsApp data stored in the Android file system and neither did they provide the detailed table structure of the database files used by WhatsApp in Android. The folder 'databases' was identified as storing the most valuable information in order to reconstruct the communication history in WhatsApp, which were the two main database files: msgstore.db and wa.db that contain all the exchanged attachments, such as images, video and contact cards. Similarly to the iOS findings, the communication activity is stored in the database called msgstore.db and the WhatsApp contact information is stored in the wa.db database file. However, compared to the WhatsApp database table structure in iOS, the Android ones contained far less tables.

## 5  Conclusion and Future work

This research was mainly motivated by the fact that despite the fact that IM applications are a ubiquitous communication tool nowadays, there was in-depth research on the topic of evidence collection from IM services. In this paper, we take into account the fact that criminal activities are taking place over IM communications in an increased frequency, the problem statement was created around providing answers on how evidence can be collected when IM is used. To that end, this research has provided elaborate answers on the types of artefacts that can be identified by the four most prevalent IM applications. Specifically, the following should now be well known for each of the investigated IM applications such as (i) the types of artefacts that can be collected in both iOS and Android devices, (ii) the location of the artefacts in each of the aforementioned file systems, (iii) the format of the artefacts and how to analyse them, (iv) a comprehensive description of the internals of every important application database with an explanation of what is stored in each database table for every IM application in both operating systems.

There are still a few remaining parts of the problem statement to be solved which could formulate future research work such as (i) the possibilities and limitations of using session cloning as an alternative investigation method in order to perform IM communication interception and (ii) the possibilities and limitations of retrieving information via the IM vendors.

## References

1. European Commission, "Digital Agenda for Europe - Telecoms and the Internet.": http://ec.europa.eu/digital-agenda/en/telecoms-and-internet.
2. ITU (International Telecommunication Union), "Global ICT developments, 2001-2013," http://www.itu.int/en/ITU-D/Statistics/Documents/statistics/2012/stat_page_all_charts.xls..
3. Portio Research, "Portio Research Mobile Factbook 2013," 2013.
4. Eurostat, "Internet use in households and by individuals in 2012," 2012.
5. UNODC, "Comprehensive Study on Cybercrime," 2013.
6. McAfee, "Hackers Using IM for Cyber Crime," 2013: http://home.mcafee.com/advicecenter/?id=ad_cybercrime_huifcc.
7. The Register, "Italian crooks use Skype to frustrate wiretaps," 2009.
8. Europol, "Threat Assessment - Italian organised crime," 2013.
9. M. Simon and J. Slay, "Voice over IP: Forensic computing implications," 2006.

10. M. Simon and J. Slay, "Investigating Modern Communication Technologies: The effect of Internet-based Communication Technologies on the Investigation Process," The journal of digital forensics, security and law, vol. 6, no. 4, pp. 35–62, 2011.
11. M. Kiley, S. Dankner, and M. Rogers, "Forensic Analysis of Volatile Instant Messaging," in in Advances in Digital Forensics IV, vol. 285, 2008, pp. 129–138.
12. M. Simon and J. Slay, "Recovery of Skype Application Activity Data from Physical Memory," 2010 International Conference on Availability, Reliability and Security, pp. 283–288, Feb. 2010.
13. T. Vidas, C. Zhang, and N. Christin, "Toward a general collection methodology for Android devices," Digital Investigation, vol. 8, pp. S14–S24, Aug. 2011.
14. K. Alghafli, A. Jones, and T. Martin, "Guidelines for the digital forensic processing of smartphones," in 9th Australian Digital Forensics Conference, 2011, no. 1, pp. 1–8.
15. C. Carpene, "Looking to iPhone backup files for evidence extraction," in Proceedings of the 9th Australian Digital Forensics Conference, 2011, pp. 16–32.
16. Y.-C. Tso, S.-J. Wang, C.-T. Huang, and W.-J. Wang, "iPhone social networking for evidence investigations using iTunes forensics," in Proceedings of the 6th International Conference on Ubiquitous Information Management and Communication - ICUIMC 12, 2012, p. 1.
17. E. R. W. Sebastian Schrittwieser, Peter Fruehwirt, Peter Kieseberg, Manuel Leithner, Martin Mulazzani, Markus Huber, "Guess Who's Texting You? Evaluating the Security of Smartphone Messaging Applications," in In proceeding of: Network and Distributed System Security Symposium (NDSS 2012), 2012.
18. H. Chu, S. Yang, S. Wang, and J. Park, "The Partial Digital Evidence Disclosure in Respect to the Instant Messaging Embedded in Viber Application Regarding an Android Smart Phone," in Proceedings of the 4th FTRA International Conference on Information Technology Convergence and Services (ITCS-12), 2012, pp. 171–178.
19. A. Mahajan, M. Dahiya, and H. Sanghvi, "Forensic Analysis of Instant Messenger Applications on Android Devices," International Journal of Computer Applications, vol. 68, no. 8, pp. 38–44, 2013.
20. C. Leung and Y. Chan, "Network forensic on encrypted peer-to-peer voip traffics and the detection, blocking, and prioritization of skype traffics," Enabling Technologies: Infrastructure for …, pp. 1–6, 2007.
21. L-M. Aouad, M. Tahar Kechadi, Justin Trentesaux, N-A. Le-Khac: An Open Framework for Smartphone Evidence Acquisition. IFIP Int. Conf. Digital Forensics 2012: 159-166.
22. Xyologic Mobile Analysis GmbH, "IM application usage statistics." http://xyo.net/.
23. Wall Street Journal, "WhatsApp Surpasses 250 Million Active Users.": http://blogs.wsj.com/digits/2013/06/20/whatsapp-surpasses-250-million-active-users/.
24. The Verge, "Viber expands to PC and Mac as competitors preach 'mobile only':www.theverge.com/2013/5/7/4305350/viber-pc-and-mac-apps-200-million-users..
25. Microsoft, "Earnings Release FY13 Q1.": http://www.microsoft.com /investor/EarningsAndFinancials/Earnings/PressReleaseAndWebcast/FY13/Q1/default.aspx.
26. Digital Trends, "Messaging app Tango steps into social network status with new photo filters and 100M users.".
27. "EDPS Glossary - Traffic Data." .
28. "Location Data - ICO." . Available: http://www.ico.org.uk/for_organisations/ privacy_and_electronic_communications/ the_guide/location_data.
29. "Skype Forensic Artifacts.": http://forensicartifacts.com/2010/08/skype/.
30. Chu, Deng, and Chao, "The digital forensics of portable electronic communication devices based on a Skype IM session of a pocket PC for NGC," Wireless Communications Mobile Computing, vol. 11, pp. 211–225, 2011.